\documentclass{aa}
\include{psfig}

\newcommand{\rms}{{\rm rms}}
\newcommand{\cd}{\cdot}
\newcommand{\ti}{\times}
\newcommand{\tii}{\!\ti\!}

\newcommand{\na}{\nabla}
\newcommand{\pa}{\partial}
\newcommand{\mb}[1]{\mbox{#1}}
\newcommand{\cu}{{\rm curl}}
\newcommand{\emf}{{\cal E}}
\newcommand{\ov}[1]{\overline{#1}}

\begin{document}
\title{Magnetoconvection and dynamo coefficients:}
\subtitle{II. Field-direction dependent pumping of magnetic field}
\author{M. Ossendrijver\inst{1} \and M. Stix\inst{1}
\and A. Brandenburg\inst{2}
\and G. R\"udiger\inst{3}}
\institute{Kiepenheuer-Institut f\"ur Sonnenphysik, Sch\"oneckstra{\ss}e 6, 79104 Freiburg, Germany
\and Nordita, Blegdamsvej 17, DK-2100 Copenhagen \O, Denmark
\and Astrophysikalisches Institut Potsdam, An der Sternwarte 16, 14482 Potsdam, Germany}
\date{Today is \today.\,\, Received; accepted}

\abstract{
We study the pumping of magnetic flux in three-dimensional compressible magnetoconvection in the context 
of stellar dynamos. The simulation domain represents a rectangular section from the lower part of a stellar 
convection zone plus the underlying stably stratified layer, with a total depth of up to five pressure
scale heights. Once convection has attained a statistically stationary state, a magnetic field is introduced.
The magnetic field is subsequently modified by the convective motions, and the resulting pumping effects are 
isolated by calculating various coefficients of the expansion of the electromotive force, $\ov{\vec{u}\ti\vec{b}}$, 
in terms of components of the mean magnetic field. The dependence of the pumping effects on rotation, latitude and 
other parameters is studied. First numerical evidence is found for the existence of pumping effects in the horizontal 
directions, unless the rotation axis coincides with the vertical axis, as is the case on the poles. 
Evidence is found that the pumping effects act differently on different components of the mean magnetic field. 
Latitudinal pumping is mainly equatorward for toroidal field, and can be poleward for poloidal field. 
Longitudinal pumping is mainly retrograde for the radial field but prograde for the latitudinal field. 
The pumping effect in the vertical direction is found to be dominated by the 
{\em diamagnetic effect}, equivalent to a predominating downward advection with a maximum speed in the turbulent 
case of about $10\%$ of the rms convective velocity. Where possible, an attempt is made to identify 
the physical origin of the effect. Finally, some consequences of the results for stellar dynamos are discussed.
\keywords{Magnetohydrodynamics (MHD) -- convection -- stars: magnetic fields -- Sun: magnetic fields}
}

\maketitle

\section{Introduction}
Magnetic pumping refers to transport of magnetic fields in convective layers which does not result from
bulk motion. The existence of magnetic pumping has been known for several decades now.
Drobyshevski \& Yuferev (\cite{drobyshevski74}) demonstrated in a numerical investigation that
magnetic flux in a convective layer is expelled downward if the downflows form connected lanes that separate
isolated upflowing regions, an effect known since as topological pumping (cf. also Moffatt \cite{moffatt78},
p. 70; Petrovay \cite{petrovay91}). A different but related effect is {\it turbulent} pumping. For the two-dimensional case,
Drobyshevski (\cite{drobyshevski77}) showed that a poloidal magnetic field is transported against the gradient 
of the density. Analytical investigations based on the first order smoothing approximation (FOSA) suggest that 
turbulent pumping also occurs in the presence of a gradient in the intensity of the turbulence, due to the 
diamagnetic effect (R\"adler \cite{raedler68}, Roberts \& Soward \cite{roberts75}, Krause \& R\"adler
\cite{krause80}, Kichatinov \& R\"udiger \cite{kichatinov92}), or as a result of rotationally induced 
anisotropies (Kichatinov \cite{kichatinov91}), or through magnetic buoyancy of small-scale fields 
(Kichatinov \& Pipin \cite{kichatinov93}). Numerical simulations of stratified convection showed that
dynamo-generated magnetic fields are pumped downward (Nordlund et al. \cite{nordlund92},
Brandenburg et al. \cite{brandenburg96}). Tobias et al. (\cite{tobias98}, \cite{tobias01}) and
Dorch \& Nordlund (\cite{dorch01}) performed detailed numerical experiments of magnetic pumping in box 
simulations of magnetoconvection. They did not, however, measure the pumping velocity which plays a role 
in mean-field electrodynamics.

Although in most investigations the possible significance of magnetic pumping for stellar dynamo action
is stressed, it was in fact rarely incorporated in  models. This is true partly because early mean-field
solar dynamo models did not appear to `need' any pumping effect since they were able to reproduce the
large-scale solar magnetic field rather well using the simplest assumption that there are just
alpha effect and differential rotation (for the alpha
effect see, e.g., Parker \cite{parker55}, or Krause \& R\"adler \cite{krause80}). Hence the pumping effect,
along with many other dynamo coefficients, was thought to be an unnecessary complication. First indications
for the usefulness of turbulent pumping came from an $\alpha\Omega$-dynamo simulation of Brandenburg et al.
(\cite{brandenburg92}), where downward pumping of the mean magnetic field was found to favour an equatorward
migration of magnetic field in the presence of mainly latitudinal differential rotation.

A major problem in solar dynamo theory was the storage of magnetic
flux for sufficiently long times (see, e.g., Moreno-Insertis et al. \cite{morenoinsertis92}).
It was believed that magnetic buoyancy would remove magnetic
field quickly from the convection zone. However, numerical simulations of
dynamo action in turbulent stratified convection indicated quite clearly
that the effects of magnetic buoyancy can be offset by the downward pumping
(Brandenburg \& Tuominen \cite{brandenburg91}, Nordlund et al. \cite{nordlund92}, Brandenburg et al.
\cite{brandenburg96}, Sect. 4.4).
Thus the vertical pumping effect may be important because sufficient amplification
of the toroidal magnetic fields generated by the solar dynamo can be achieved only in or near the stably
stratified overshoot layer below the convection zone. 

Apart from the vertical direction, pumping effects can also exist in other directions. So far though, non-radial 
pumping effects have not received much attention, although they have interesting consequences for stellar dynamos, 
as is discussed further below. Their existence has been predicted on the basis of symmetry 
considerations within the framework of mean-field electrodynamics (Krause \& R\"adler \cite{krause80}), and 
some analytical results based on FOSA are presented in Kichatinov (\cite{kichatinov91}) and 
R\"udiger \& Kichatinov (\cite{ruediger93}). No numerical studies of non-radial pumping effects have previously
been carried out, as far as we know. The present investigation provides the first numerical evidence for their 
existence, as well as quantitative results for the corresponding transport coefficients.

The study of pumping effects is part of an effort to calculate dynamo coefficients
through numerical simulations. This work is motivated by the idea that progress in understanding
stellar dynamo action can be achieved only through a combination of approaches which include numerical
simulation as well as mean-field modelling. On the one hand, fully global MHD calculations of stellar
dynamos have not been successful so far because computer capacities are insufficient to
capture the parameter regime of stellar convection zones. Local MHD simulations of small subvolumes
offer the best possibility to approach solar parameters.
On the other hand, observations of the large-scale solar magnetic field (solar cycle, butterfly
diagram, Hale's polarity laws) suggest that some form of mean field description is possible.
Local numerical investigations of aspects of stellar dynamos may provide useful input information
for the construction of mean-field models. Also, the study of dynamo coefficients
can be a powerful diagnostic tool to isolate from numerical simulations the various effects that play a role in
stellar dynamos and to quantify them. In order to understand the role of dynamo coefficients, consider
the averaged MHD induction equation:
\begin{equation}
\frac{\pa\ov{\vec{B}}}{\pa t}=\cu\,\big(\ov{\vec{U}}\tii\ov{\vec{B}}+
   \vec{\emf}-\eta\,\cu\,\ov{\vec{B}}\big)\,,
\end{equation}
Here $\ov{\vec{B}}$ is the mean magnetic field, $\ov{\vec{U}}$ is the mean flow, $\eta$ is the magnetic
diffusivity, and $\vec{\emf}=\ov{\vec{u}\ti\vec{b}}$ is the {\em electromotive force} (EMF) which depends on
$\vec{u}=\vec{U}-\ov{\vec{U}}$, the fluctuating component of the flow, and $\vec{b}=\vec{B}-\ov{\vec{B}}$,
the fluctuating component of the magnetic field. Mean quantities are either ensemble averages or
azimuthal averages. In the present case of a box with periodic boundary conditions in the horizontal
directions (see below), horizontal averages can be used instead, and in practice, there is additional 
temporal averaging. The dynamo coefficients appear in an expansion of the
electromotive force in terms of spatial derivatives of the mean magnetic field. In Cartesian coordinates,
and assuming the summation convention, this may be written as
\begin{equation}
\emf_i=\alpha_{ij}\ov{B}_j+\beta_{ijk} \na_j\ov{B}_k+\cdot\cdot\cdot\,. \label{emf}
\end{equation}
The coefficients in the expansion, of which only the first two are shown,
are pseudo tensors of increasing rank. In what follows, we consider only the first term, i.e. we
assume that the mean magnetic field in the simulation domain is approximately independent of spatial coordinates.
Commonly one separates the $\alpha$-tensor into symmetric and antisymmetric parts, i.e.
\begin{equation}
\alpha^{\rm S}_{ij}=(\alpha_{ij}+\alpha_{ji})/2;\hspace{1cm}
\alpha^{\rm A}_{ij}=(\alpha_{ij}-\alpha_{ji})/2. \label{as}
\end{equation}
The latter gives rise to an advection term, $\alpha^{\rm A}\cd\ov{\vec{B}}=\vec{\gamma}\ti\ov{\vec{B}}$, where
\begin{equation}
\vec{\gamma}\!=\!-\pmatrix{\alpha^{\rm A}_{yz}\cr\!\alpha^{\rm A}_{zx}\!\cr\alpha^{\rm A}_{xy}}.
\label{gammagen}
\end{equation}
This term describes the pumping of all components of the magnetic field with equal speed, and we shall refer
to it as the `general pumping effect'. Note that the pumping effect as it is used here refers to the zeroth-order
terms in the expansion of the EMF only. Turbulent diffusion ($\beta$) and higher order contributions
can also give rise to transport phenomena of the mean magnetic field, but they cannot be expressed by
advection terms and are to be treated separately.
One may also ask whether there are circumstances under which different components of the magnetic field have
different advection speeds and directions. A necessary condition that must be fulfilled when addressing
this question is that the magnetic field can be decomposed in divergence-free components. Only then can one
write down separate equations for the magnetic flux of each field component through a surface moving with a velocity
$\vec{w}$, say, thereby allowing one to distinguish advection from other processes.
This holds, for example, for a
decomposition using toroidal and poloidal components (Kichatinov \cite{kichatinov91}).
But unlike Alfv\'en's theorem in ideal MHD (where $\vec{w}=\vec{U}$ for all divergence-free components),
it is generally impossible in mean-field electrodynamics to define a
moving surface in which flux is conserved. That is due to the presence of the alpha effect and turbulent
diffusion. Essentially, what can be achieved is to define a velocity that optimizes flux
conservation for a given divergence-free component. For example, in the case of $\ov{\vec{B}}_{\rm tor}$,
one sets $\vec{w}_{\rm tor}=\ov{\vec{U}}+\vec{\gamma}^{\rm (tor)}$, where $\vec{\gamma}^{\rm (tor)}$ denotes the
pumping velocity of the toroidal field (even though it is itself poloidal).

As a result of the averaging over the horizontal coordinates, our mean magnetic field is of 
the form $\ov{\vec{B}}=[\ov{B}_x(z),\ov{B}_y(z),\ov{B}_z]$, so that all three Cartesian components are 
divergence free. If one interprets this field as the mean field of an axisymmetric spherical dynamo, then
$\ov{B}_x\vec{e}_x+\ov{B}_z\vec{e}_z$ corresponds to the poloidal component and $\ov{B}_y\vec{e}_y$ to
the toroidal. The meaning of the tensor components $\alpha_{ij}$ may be grasped most easily by writing
\begin{equation}
\vec{\emf}={\bf\alpha}^{\rm D}\cd\ov{\vec{B}} +\vec{\gamma}^{(x)}\ti\ov{\vec{B}}_x+
 \vec{\gamma}^{(y)}\ti\ov{\vec{B}}_y+\vec{\gamma}^{(z)}\ti\ov{\vec{B}}_z\,,\label{emf2}
\end{equation}
where $\ov{\vec{B}}_x=\ov{B}_x\vec{e}_x$, etc. Here, ${\bf\alpha}^{\rm D}$ denotes the diagonal part of 
${\bf\alpha}$, which is responsible for maintaining a large-scale magnetic field in solar-type stars.
The off-diagonal terms of the $\alpha$ tensor represent various contributions to the pumping effect.
As shown by Eq. (\ref{emf2}), they are equivalent to an effective advection of the mean magnetic field, where
\begin{eqnarray}
\vec{\gamma}^{(x)}\!&=&\!\vec{\gamma}+\pmatrix{0\cr\!-\alpha_{xz}^{\rm S}\!\cr\alpha_{xy}^{\rm S}},\hspace{4mm}
\vec{\gamma}^{(y)}\!=\!\vec{\gamma}+\pmatrix{\alpha_{yz}^{\rm S}\cr0\cr\!-\alpha_{xy}^{\rm S}\!},\;\nonumber\\[3mm]
\vec{\gamma}^{(z)}\!&=&\!\vec{\gamma}+\pmatrix{\!-\alpha_{yz}^{\rm S}\!\cr\alpha_{xz}^{\rm S}\cr0} \label{up}
\end{eqnarray}
are the pumping velocities for the respective Cartesian components of the mean field. This representation  
shows that the pumping velocity can, in principle, be different for each component, as was pointed out by 
Kichatinov (\cite{kichatinov91}), and that the differences are equal to the off-diagonal, symmetric components 
of the $\alpha$-tensor. 

In general, the nine components $\alpha_{ij}$ are independent. At the poles however, the rotation
axis coincides with the direction of gravity, resulting in isotropy in the horizontal plane.
Since the only remaining preferential direction is the vertical one, it follows from elementary symmetry
considerations that the pumping velocity cannot have horizontal components, i.e.
$\alpha_{xz}=\alpha_{zx}=\alpha_{yz}=\alpha_{zy}=0$, and that the pumping effect acts in identical ways on
$\ov{B}_x$ and $\ov{B}_y$, i.e. $\alpha_{xy}=-\alpha_{yx}$. By the same token $\alpha_{xx}=\alpha_{yy}$.
Hence at the poles the EMF is given by $\vec{\emf}=\alpha_{xx}\ov{\vec{B}}_{\rm H}+\alpha_{zz}\ov{\vec{B}}_z
+\vec{\gamma}\ti\ov{\vec{B}}$, where $\ov{\vec{B}}_{\rm H}=[\ov{B}_x,\ov{B}_y,0]$ is the horizontal mean
field and $\vec{\gamma}=-\alpha_{xy}\vec{e}_z$ is the pumping velocity. At the equator, similar arguments
lead to the conclusion that the pumping velocity in the $x$-direction vanishes, i.e. $\alpha_{yz}=\alpha_{zy}=0$,
and that ${\bf \alpha}^{\rm D}=0$. These same components are antisymmetric with respect to the
equatorial plane, while the other ones are symmetric.

The remainder of the paper is structured as follows. After introducing the model (Sect.~2), we explain the method 
by which the dynamo coefficients are calculated (Sect.~3). In Sect.~4 we present results on the pumping effects, their 
dependence on various parameters, and we make a qualitative comparison with analytical results,
where they are available. Briefly, some results on the diagonal alpha coefficients will be mentioned. Finally, the 
results and their possible implications for stellar dynamos are discussed (Sect.~5).

\section{Model}
\begin{figure}
\centerline{\psfig{file=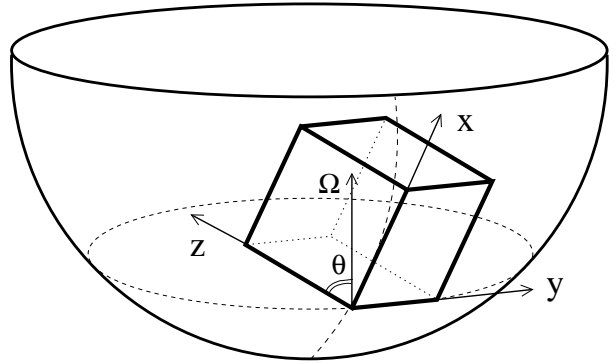,width=8cm}}
\caption{Geometry of the simulation domain.}
\label{fig1}
\end{figure}

For a detailed description of the model and for the full equations and boundary conditions, the reader is
referred to Ossendrijver et al. (\cite{ossendrijver01}; hereafter cited as Paper I). The simulation domain
consists of a rectangular box which is defined on a Cartesian grid, with $x$ and $y$ denoting the two
horizontal coordinates (corresponding to latitude and longitude in spherical coordinates), and $z$ denoting
depth, respectively (Fig.~\ref{fig1}). Gravity is directed in the positive $z$-direction.
The box rotates about an axis in the $xz$-plane,
$\vec{\Omega}=\Omega\,(\vec{e}_x\sin\theta+\vec{e}_z\cos\theta)$, where $\theta$ is the angle between the
rotation vector and the $z$-axis, or the angular distance to the south pole.

From top to bottom, the simulation domain consists of a thin cooling layer, a convectively unstable layer,
and a stably stratified layer with overshooting convection. The consecutive boundaries of the layers are
located at depths $z_1$, $z_2$, $z_3$ and $z_4$, where $z_2=0$ and $z_3=d$ in all cases. The horizontal
extent of the box, $L\ti L$, is chosen such that there are always at least a few convective cells within
the box. The governing equations are those describing magnetic induction, mass
continuity, and the balance of momentum and energy. Solutions are governed by the following independent
dimensionless parameters. The Prandtl number and the magnetic Prandtl number are defined as
\begin{equation}
\mb{Pr}=\nu/\chi_0\,, \hspace{1cm} \mb{Pm}=\nu/\eta\,, \label{pr}
\end{equation}
where $\chi_0=\kappa_2/\gamma_{\rm ad}\rho_0$ denotes a reference value of the radiative
diffusivity for region 2. In order to avoid possible confusion with the pumping effect we write for
the ratio of specific heats $\gamma_{\rm ad}=5/3$. The parameter $\xi_0$ determines the internal energy
at the top, $e_0$, and hence the pressure scale height there, i.e.
\begin{equation}
\xi_0=(\gamma_{\rm ad}-1)\,e_0/gd\,.
\end{equation}
The initial thermal structure of the three regions is characterized by polytropic indices
$m_i=(1-\na_i)/\na_i$, where $\na_i=(d\ln T/d\ln p)_i$ denotes the radiative temperature gradient
($i=1,2,3$). The adiabatic temperature gradient is $\na_{\rm ad}=(\gamma_{\rm ad}-1)/\gamma_{\rm ad}$.
A measure for instability is provided by the (initial) superadiabaticity,
$\delta_i=\na_i-\na_{\rm ad}$, which is positive in an unstably stratified medium. The reference Rayleigh
number is defined as
\begin{equation}
\mb{Ra}=\frac{d^4 g\delta_2}{\nu\chi_0 H_{ph}}\,, \label{ra}
\end{equation}
where $H_{ph}=\xi_0 d+0.5 d/(m_2+1)$ is the pressure scale height in the middle of the
unstable layer, and the parameters $\delta_2$ and $H_{ph}$ refer to the initial stratification.
The Taylor number,
\begin{equation}
\mb{Ta}=(2\Omega d^2/\nu)^2\,,
\end{equation}
measures the importance of rotation relative to viscous dissipation. Finally, $\sigma_0$
represents the rate at which internal energy adjusts to the specified reference value $e_0$.
The remaining parameters are secondary. The Coriolis number, or inverse Rossby number, measures
the importance of the Coriolis force:
\begin{equation}
\mb{Co}=2\Omega \tau\,.
\end{equation}
Here $\tau=d/u_{\rms}$ is an estimate of the correlation time, and $u_{\rm rms}$ is the rms velocity defined
in a suitable way (e.g., by averaging over time and over the bulk of the convection zone in the box).
In the initial state, the
radiative energy flux, $F_{{\rm rad},z}=-\kappa de/dz$, is assumed to be constant throughout the domain. This
determines the radiative conductivities in the three regions
according to $\kappa_i/\kappa_2=(m_i+1)/(m_2+1)$. In order to obtain a continuous function, $\kappa$
is smoothed across thin intermediate layers between the three regions.
The initial stratification, obtained from a simplified mixing length model, is calculated with the
constraint $\rho(z_3)=\rho_0$. Kinetic and magnetic Reynolds numbers are defined as
\begin{equation}
\mb{Re}=u_{\rm rms}d/\nu\,, \hspace{1cm} \mb{Rm}=u_{\rm rms}d/\eta\,.
\end{equation}
Dimensionless units are introduced by setting $d=\rho_0=g=\mu_0=1$, where $\mu_0$ is the vacuum
permeability.

\begin{table}[htb]
\caption{Parameters. In all runs, $z_1=-0.07$, $z_2=0$, $z_3=1$, $z_4=1.93$, $L=4$, $\xi_0=0.1$, $m_1=0$,
         $m_2=1$ (i.e., $\delta_2=0.1$), $m_3=3$, $B_0=0.001$, $\mb{Pr}=0.4$ and $\mb{Pm}=1$ 
         (i.e., $\mb{Rm}=\mb{Re}$). From these parameters, and Ra, $\nu$ is derived using Eq.~(\protect{\ref{ra}}). 
         The quantities $u_{\rms}$, $\mb{Ma}$, $\mb{Re}$ and $\mb{Co}$ are averages over
         time and over the unstable layer; $\Delta\ln\ov{p}_{\rm CZ}$ is the number of pressure scale heights
         accross the unstable layer. The unit of length is $d$, that of $u_{\rms}$ is $\sqrt{dg}$,
         and that of $\nu$ is $\sqrt{d^3 g}$.}
\begin{center}
\begin{tabular}{cccccc} \hline\noalign{\smallskip}
series        & A           & B           & C             & D 	   & E           \\
\noalign{\smallskip}\hline\noalign{\smallskip}
grid          & $80^3$      & $80^3$      & $160^2\ti 80$ & $50^3$ & $50^3$      \\
Ra 	      & $10^6$      & $10^6$      & $4\cd 10^6$	  & $10^4$ & $10^4$      \\
Ta            & $4\cd 10^5$ & $4\cd 10^4$ & $4\cd 10^5$	  & $10^2$ & $5\cd 10^3$ \\
$\nu\cd 10^3$ & $0.34$      & $0.34$      & $0.17$        & $3.4$  & $3.4$        \\
$u_{\rms}$    & $0.090$     & $0.086$     & $0.089$       & $0.11$ & $0.10$      \\
Co            & $2.4$       & $0.75$      & $1.2$         & $0.3$  & $2$         \\
Ma            & $0.09$      & $0.09$	  & $0.11$        & $0.12$ & $0.12$      \\
Re            & $260$       & $260$       & $524$	  & $32$   & $32$        \\
$\Delta\ln\ov{p}_{\rm CZ}$ & $3.7$ & $3.7$ & $3.8$        & $2.8$  & $2.8$       \\
\noalign{\smallskip}\hline
\end{tabular}
\end{center}
\label{partable}
\end{table}

\section{Method}
Initially, a simulation is performed without a magnetic field until convection has attained a statistically
stationary state. Then at most three numerical experiments are done with imposed weak homogeneous magnetic
fields in orthogonal directions. For each run, the EMF and the mean magnetic field are calculated, and
by combining the results, the nine coefficients $\alpha_{ij}$ are determined through the relation
$\vec{\emf}={\bf\alpha}\cd\ov{\vec{B}}$. A necessary condition for the results to be valid is that the
statistical properties of the flow are independent of the imposed magnetic field, i.e. that no {\it additional}
anisotropies are introduced. This is ensured because during the simulation, the field strength remains
small compared to the equipartition field strength, so that the Lorentz force is weak. 

\begin{figure*}
\centerline{\psfig{file=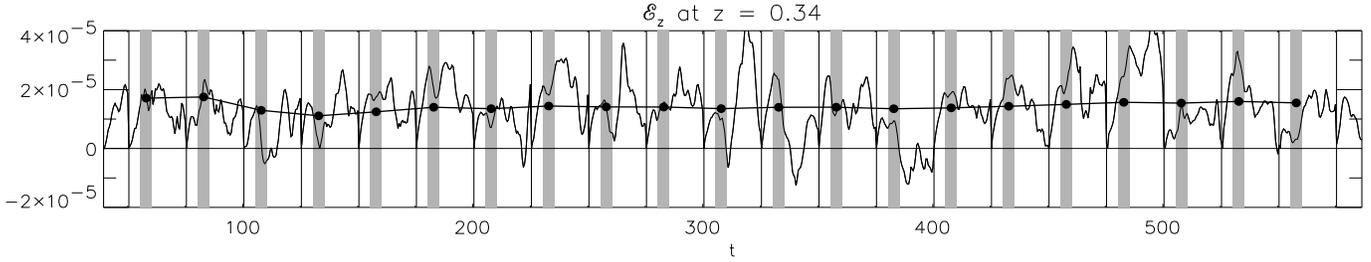,width=18.4cm}}
\caption{Averaging procedure as applied to run A45X. The thin curve is $\emf_z$ at $z=0.34$, averaged over 
the horizontal coordinates only, as a function of time (in units of $(d/g)^{1/2}$). 
Vertical lines demark the times at which the magnetic field is 
initialised ($T=25$), and gray bands denote the optimal subinterval, $[t_1,t_2]$, used for the temporal averaging. 
The resulting cumulative average, shown as dots connected by thick straight lines, converges after typically 
a few initialisations. For the parameters of this run, see Table~\protect{\ref{partable}}.}
\label{fig2a}
\end{figure*}

\begin{figure}
\centerline{\psfig{file=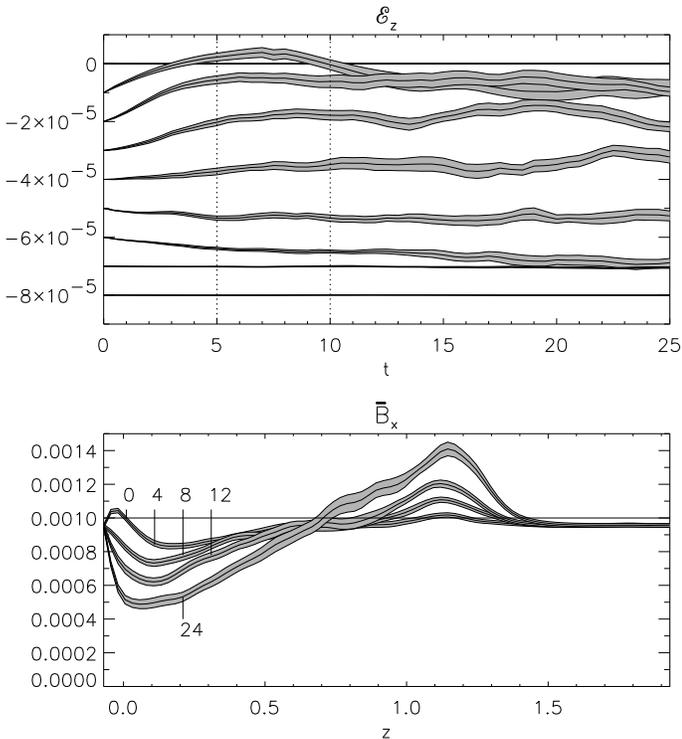,width=9.2cm}}
\caption{Averaging procedure as applied to run A45X. Top:  $\emf_z$ as a function of time at 9 equidistant 
locations between $z=-0.07$ (top curve) and $z=1.55$ (bottom curve). A vertical shift has been applied for 
clarity; all curves begin at the origin. The dotted vertical lines demark the subinterval $[t_1,t_2]$. 
For the third curve ($z=0.34$), the average over this interval corresponds to the last point in
Fig.~\protect{\ref{fig2a}}. Bottom: $\ov{B}_x$ as a function of depth (in units of $d$) at $t=0,4,8,12,24$. 
At $t=0$, the magnetic field is set to $B_x=B_0=0.001$. The averages in the lower two figures are 
calculated from Eq. (\protect{\ref{avslice}}), with $T=25$ and $N=21$. The vertical extent of 
the shaded areas is $2\sigma_{\rm mean}$ (see text). For the parameters of this run, see 
Table~\protect{\ref{partable}}.}
\label{fig2}
\end{figure}

As was pointed out in Paper I, the determination of dynamo coefficients with sufficient accuracy requires
both spatial averaging and temporal averaging over at least several turnover times. Compared to Paper I, we
employ an improved method that gives more accurate results than can be obtained by performing an average over
the horizontal coordinates plus a simple time average. The method aims to minimize
contributions to the EMF (\ref{emf}) due to first and higher order derivatives of the mean magnetic field.
Such contributions can arise because with the exception of $\ov{B}_z$, mean quantities, defined as horizontal 
averages, can have a dependence on $z$. Thus, a few turnover times after the introduction of a homogeneous 
magnetic field, the mean field is characterized by rather steep
gradients at several locations. The only exception is $\ov{B}_z$, which by virtue of the
periodic boundary conditions in the horizontal directions must remain unchanged.
In the ideal situation where the $z$-derivatives of the mean field vanish, the
only contributing coefficients are $\alpha_{ij}$, and they can be calculated exactly by performing a
series of (at most) three runs using imposed magnetic fields with orthogonal orientations.
Note that for $\theta=0\degr$ or $180\degr$ a pumping effect exists only in the $z$-direction; it can be
measured from a single run with an imposed  field in the $x$ or $y$ directions.
In the presence of significant nonvanishing gradients on the other hand, one
could account for the higher-order contributions by doing more experiments using imposed
magnetic fields that have gradients and calculating the corresponding dynamo coefficients. 
Even from one experiment one can calculate the additional diffusive contributions if the mean field is varying 
in space and time. This allowed Brandenburg \& Sokoloff (\cite{brandenburg02}) to estimate all components of 
the $\alpha$ and $\beta$ tensors (Eq.~\ref{emf}) from local simulations of accretion discs which feature 
migratory dynamo cycles. 

Instead of following this very cumbersome recipe,
we have proceeded by periodically resetting the magnetic field in the entire box to the same initial
homogeneous field at times $t_0+iT$, where $i=0,1,\cd\cd\cd,N$, thereby allowing the field to evolve for
only a limited duration $T$. In this manner, a sequence of $N$ time slices is obtained which, to good
approximation, are independent realisations. By averaging a quantity, say $A$, over the horizontal
coordinates and over the time slices according to
\begin{equation}
\ov{A}(z,t)=\frac{1}{NL^2} \sum_{i=0}^{N-1}\!\int\!\!\!\int \!dx\,dy\,A(x,y,z,t_0+t+iT), \label{avslice}
\end{equation}
where $0\leq t<T$, one obtains something equivalent to an ensemble
average of a time slice. The `error' in $\ov{A}(z,t)$ is estimated from the standard
deviation, $\sigma$, obtained from all the time slices according to $\sigma_{\rm mean}=\sigma/\sqrt{N}$.
In order to obtain significant results $N$ must exceed a minimal value which depends on several factors 
such as the magnitude of the coefficient that one tries to measure, the amount of spatial averaging, 
and the level of turbulence; for the present runs, it was found that $N$ should be at least $5$.

The most delicate part of the method consists of identifying an optimal subinterval within the mean 
time slice, say $t_1\leq t\leq t_2$,
that provides the most reliable estimate of the dynamo coefficients. The best estimate will be defined as
the time average over this subinterval. Two conditions must ideally be fulfilled. First, the EMF must have
attained the equilibrium value given by Eq. (\ref{emf}). This condition is not trivially fulfilled, because
$\vec{\emf}=\ov{\vec{u}\ti\vec{b}}=0$ at times $t_0+iT$, in apparent contradiction with Eq. (\ref{emf}).
Obviously the paradox arises because Eq. (\ref{emf}) is not valid instantaneously from the moment when a
homogeneous magnetic field is introduced, since it takes a while for the magnetic fluctuations to develop.
Thus $t_1$ must be chosen sufficiently late.
Secondly, $t_2$ must be chosen such that the gradients of the mean field
are still numerically small compared to the imposed field itself divided by, say, the thickness of the
unstable layer. If this condition is satisfied, then $\alpha_{ij}$ is the dominant contribution to the
EMF, provided that $\alpha_{ij}$ does not vanish for reasons of symmetry or other.
Both conditions can be met if the EMF evolves sufficiently more rapidly than does the
mean magnetic field. In practice, the time scales are rather similar in the present simulations, and
both conditions are at best marginally fulfilled. 

The method is illustrated in Figs.~\ref{fig2a} and \ref{fig2} for run A45X (see Table~\ref{partable} 
for the parameters of this run). Fig.~\ref{fig2a} shows the $z$-component of the EMF at a fixed depth in the box, 
defined as an average over $x$ and $y$, as a function of time. Although the temporal evolution varies strongly 
between the different initialisations, the additional averaging over the subinterval $[t_1,t_2]$ and 
over the subsequent realisations leads to a well-converged result. This may be compared with 
Fig.~2 of Cattaneo \& Hughes (\cite{cattaneo96}; henceforth referred to as CH96), which suggests that, 
for a weak initital magnetic field, convergence is achieved only after more than about $50$ turnover times. 
However, there are important differences compared with their calculation. 
First, the dynamics in CH96 are very different, because the flow results from external forcing rather 
than moderately turbulent convection, and because the Lorentz force dominates, whereas it is negligible 
in our case. We have smaller fluctuations in the horizontal averages, partly because the length scale of convection is always at 
least a few times smaller than the box size, whereas they are similar in CH96. Also, $\mbox{Re}$ 
is smaller in comparison with CH96, and therefore the level of fluctuations. 
Note that in order to compare $\mbox{Re}$ in Table~\ref{partable} with the Reynolds number, 
$U/\nu k_{\rm f}=100$, of CH96, where $k_{\rm f}$ is the wave number of the forcing, $\mbox{Re}$ should be 
multiplied with a factor of about $1/2\pi$. Secondly, the averaging in CH96 starts from an initial state where 
there is no alpha effect, which exists only after $30$ turnover times. This introduces a transient in 
the cumulative time average, which is absent in our case. 
As a result of these differences, we obtain convergence already if $N\approx 6$, even though this  
corresponds to a total averaging time, $N\cd(t_2-t_1)$, of a few turnover times only.
Fig.~\ref{fig2}, top, shows mean time slices
for the $z$-component of the EMF at different depths in the box. After an initial phase of approximately linear
growth which, depending on the depth, lasts about $4-8$ units, the EMF levels off to the equilibrium value.
After that, it continues to evolve more slowly as it adapts to the changing mean magnetic 
field in accordance with Eq.~(\ref{emf}). The lower figure shows $\ov{B}_x$ as a function of depth at 
consecutive moments.
Starting from the initial condition, the mean magnetic field evolves while developing increasing gradients,
especially near the upper and lower boundaries of the convective layer.
In the time interval $[5,10]$ the magnitude of the gradients in the bulk of the unstable layer is 
$d/dz \ln \ov{B}_x\approx 0.5$, and in the interval $[20,25]$ it is about $1$. It follows that the  
best available time interval for calculating dynamo coefficients from run A45X is roughly between $5$ and $10$.

\begin{figure}
\centerline{\psfig{file=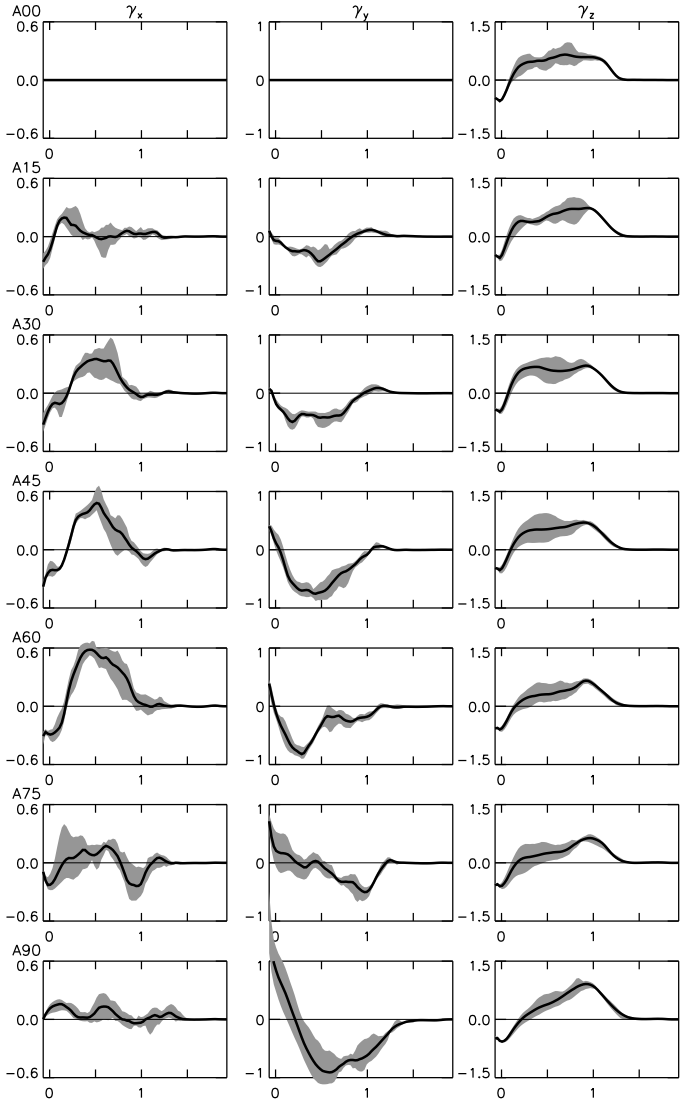,width=9.0cm}}
\caption{Pumping velocity $\vec{\gamma}$ (Eq. \protect{\ref{gammagen}}) measured in units of 
$0.01\sqrt{dg}$ as a function of depth in units of $d$ for series A. Each row, corresponding to a different 
value of $\theta$, shows the three Cartesian components of $\vec{\gamma}$ for the run indicated on the left. 
For the parameters, see Table \ref{partable}. The black curves are time averages over the optimal interval 
as defined in Section 3. Shaded areas indicate the range of values attained within the optimal time interval. 
The vertical scale is the same in each row, but differs from column to column. In series A00 only $\gamma_z$ was 
determined, while the other components of $\vec{\gamma}$ are known to vanish on symmetry grounds. On the same 
grounds one knows that $\gamma_x=0$ in series A90. This was not imposed, so that this coefficient gives an indication 
of the error of the method.} 
\label{fig3}
\end{figure}

As soon as the EMF levels off to its equilibrium
value (\ref{emf}), the mean magnetic field already begins to develop considerable gradients at some depths.
Nevertheless, the results are less contaminated with higher-order contributions than those obtained by 
performing a simple time average accross a long simulation in which the magnetic field was initialized only once.
An indication of the error in the coefficients is provided by the variation of the coefficients within the
optimal time interval as shown by the shading in Figs. \ref{fig3}--\ref{fig7}. In order to assess the 
possible difference between the results of the improved method and 
those of the less accurate method of Paper I, we have in a few cases also calculated $\alpha_{ij}$ using 
the time interval $[t_1,t_2]=[20,25]$. Although this choice of $[t_1,t_2]$ yields the correct sign and order of
magnitude, the higher-order contributions can amount to $40\%$ in some cases. All results presented below were 
therefore derived using the earliest possible interval, so that the influence of the higher-order contributions 
is minimized. As is shown in Fig.~\ref{fig2}, bottom, the gradients of the mean field can be particularly strong
near the top and the bottom of the unstable layer, so that at these locations the results should be treated with 
some caution.  To avoid misunderstandings, we emphasize that the accuracy discussed here is distinct from the error 
introduced if the time sequence is too short to average out the fluctuations.

\section{Results}
\begin{figure}
\centerline{\psfig{file=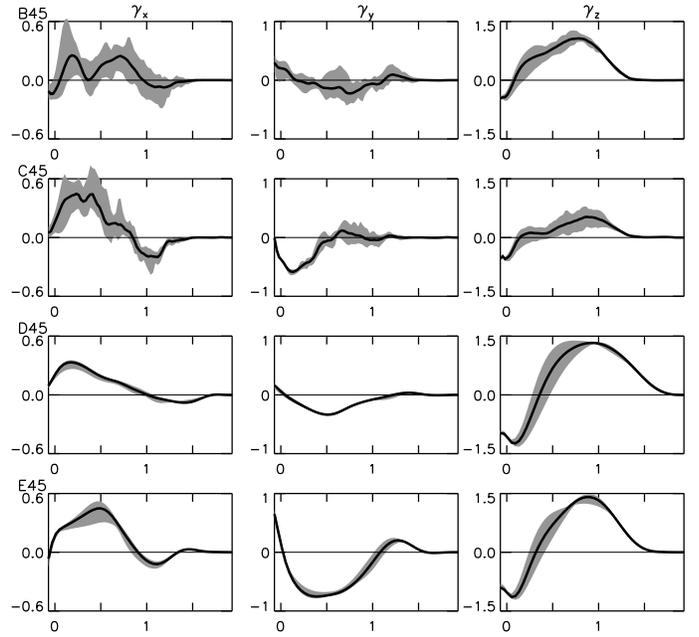,width=9.0cm}}
\caption{Pumping velocity $\vec{\gamma}$ (Eq. \protect{\ref{gammagen}}) measured in units of 
$0.01\sqrt{dg}$ as a function of depth in units of $d$ from four different runs performed at
$\theta=45\degr$ (cf. also Fig.~\protect{\ref{fig3}} and Table \ref{partable}).}
\label{fig4}
\end{figure}

\begin{figure}
\centerline{\psfig{file=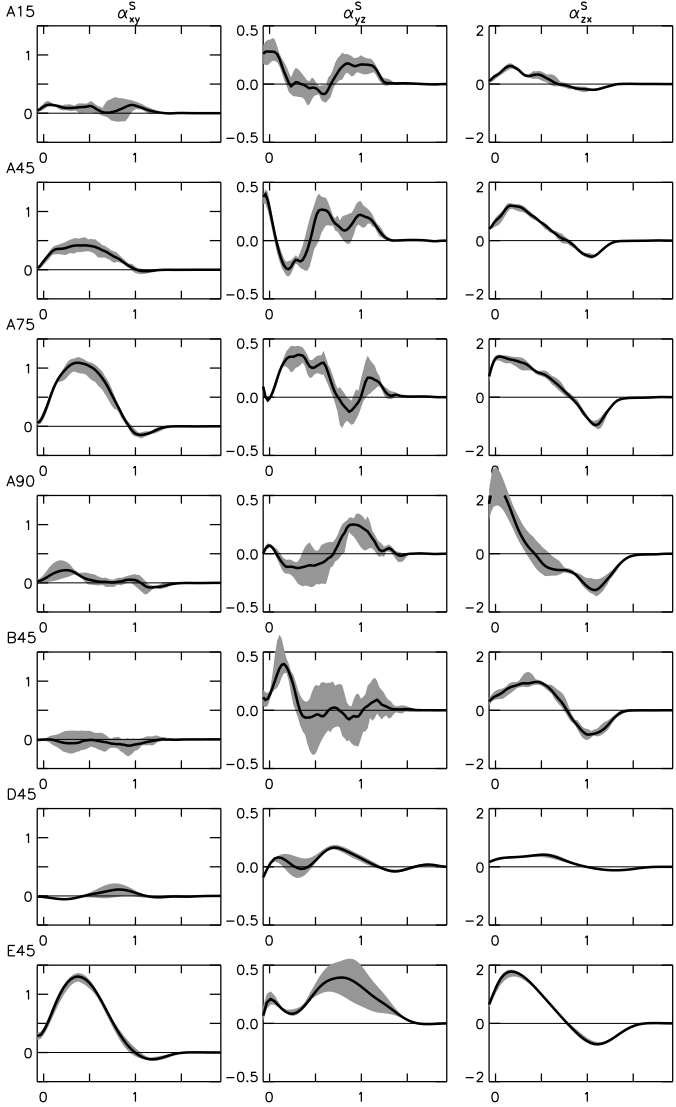,width=9.0cm}}
\caption{Off-diagonal symmetric components of the $\alpha$-tensor (Eq. \protect{\ref{as}}) measured in units of 
$0.01\sqrt{dg}$ as a function of depth in units of $d$ (cf. also Fig.~\protect{\ref{fig3}} and Table \ref{partable}).
On symmetry grounds one knows that $\alpha_{yz}^{\rm S}=0$ in series A90. This was not imposed, so that 
this coefficient gives an indication of the error of the method. In series B45 and D45, which are characterised by 
rather slow rotation, $\alpha_{xy}^{\rm S}$ is also not significant.}
\label{fig5}
\end{figure}

The numerical simulations are grouped together in several series named
A--E, each of which may contain several experiments performed at different latitudes and having
different orientations of the imposed field. Each run is referred to by a code
such as A45X, which identifies it as belonging to series A, having $\theta=45\degr$
and an imposed magnetic field in the $x$-direction. The parameters of each series are summarized in
Table \ref{partable}. They may be characterized as follows: series A is moderately turbulent and has
a rotation rate appropriate for the base of the solar convection zone. Series B has the same parameters
as A except for rotation, which is slower. Series C has the highest Rayleigh number and the most turbulent
convection. Series D has a Rayleigh number just above marginal stability. Accordingly, the flow
is laminar, and the Reynolds number is small. It has the slowest rotation of all.
Series E is identical to D, except for the rotation rate.
In series A, $\theta$ was varied between $0\degr$ and $90\degr$ with a stepsize of $15\degr$;
all other simulations were done for $\theta=45\degr$.
The optimal time interval, $[t_1,t_2]$, was determined to be $[5,10]$ for series A and B, $[4,7]$ for C, and
$[15,20]$ for D and E. The value of $N$ was in the range between $6$ and $21$. Although the 
total amount of time used in the averaging, $N\cd(t_2-t_1)$, corresponds to several turnover times only, 
the combined spatial and temporal averaging as defined by Eq. (\ref{avslice}) was in most cases 
sufficient to yield significant results. This is illustrated by the convergence of the cumulative
average shown in Fig.~\ref{fig2a}, by the well-defined shapes of the curves of $\emf_z$ and $\ov{B}_x$ 
in Fig.~\ref{fig2}, and by the narrowness of most of the shaded areas in Figs. \ref{fig3}--\ref{fig7}.  

In the following subsections we present the main results of the simulations. As will become apparent,
many of the results are new in the sense that apart from the pumping effect in the vertical direction,
none of the effects have been numerically investigated before as far as we know. Also, this is the first
numerical investigation in which pumping effects are quantified in terms of gamma coefficients. For that 
reason, and in order to compare the various pumping effects, it is worthwile to consider also the vertical 
pumping effect, although it has already received much recent attention in the literature. Results for the 
coefficients $\alpha_{ij}$ of all series are shown in Figs. \ref{fig3}--\ref{fig7} in the form of the three 
Cartesian components of $\vec{\gamma}$ and the independent elements of $\alpha_{ij}^{\rm S}$. From these, 
all components of $\alpha_{ij}$ can, if necessary, be derived using Eqs (\ref{as}) and (\ref{up}).
The black curves in Figs. \ref{fig3}--\ref{fig7} are averages of a given coefficient over the time interval
$[t_1,t_2]$. The shaded areas indicate the range of values attained at individual times within the
interval $[t_1,t_2]$, and are meant to give an error indication.
Some confirmation that the error estimates are consistent, provided that they are used conservatively,
can be seen in series A90, where $\gamma_x,\alpha_{yz}^{\rm S},\alpha_{xx},\alpha_{yy}$ and
$\alpha_{zz}$ are known to vanish on symmetry grounds (Figs.~\ref{fig3}, \ref{fig5} and \ref{fig6}). 
For these coefficients the shaded area is not always consistent with zero. This departure gives
an additional estimate of the expected error. We reiterate that results for $\alpha_{ij}$ near 
the top and the bottom of the unstable layer should be treated with caution because of the enhanced possibility 
of contamination with higher-order terms.

\subsection{Latitudinal pumping effect}
In the latitudinal direction, the simulations yield a weak net general advection, measured by $\gamma_x$,
which is equatorward ($\gamma_x>0$) in the bulk of the unstable layer and can be poleward near the upper 
surface and in the overshoot layer (Figs. \ref{fig3}--\ref{fig4}). At the pole and at the equator the effect 
vanishes; it is strongest near $\theta=60\degr$.
The symmetric contribution $\alpha_{yz}^{\rm S}$ is about as weak as $\gamma_x$, but shows little systematic
behaviour as a function of $\theta$ (Fig.~\ref{fig5}). Both increase
with increasing rotation rate, since they are rotationally induced (compare D45 and E45).
As can be inferred from the relations $\gamma^{(y)}_x=\gamma_x+\alpha_{yz}^{\rm S}$ and
$\gamma^{(z)}_x=\gamma_x-\alpha_{yz}^{\rm S}$ (Eq. \ref{up}), the $y$ and $z$ components of the mean field
have different latitudinal velocities (Fig.~\ref{fig8}). In all simulations, $\ov{B}_y$ tends to be pumped towards the equator
($\gamma^{(y)}_x>0$) in the bulk of the unstable layer. The situation with regard to the component $\ov{B}_z$ 
is less clear. It is generally pumped towards the pole ($\gamma^{(z)}_x<0$) in the lower part of the unstable layer, 
but the behaviour in the bulk of the unstable layer appears to vary with latitude. 
The average over the unstable layer turns out to be mostly positive (equatorward) for both components (Fig.~\ref{fig8}).
The lowest-order contribution to latitudinal
pumping in the expansion of the EMF presented by, e.g., Krause \& R\"adler (\cite{krause80}), p. 193, can be written 
as $\Delta\alpha_{ij}=A\Omega_i\epsilon_{jkl}g_k\Omega_l+B\Omega_j\epsilon_{ikl}g_k\Omega_l$, 
where $A$ and $B$ are true scalars, i.e. even functions of
$\cos\theta$. Thus, the latitudinal pumping is at least of second order in $\Omega$. One 
mechanism that may explain the effect is provided by the rotational anisotropies studied by
Kichatinov (\cite{kichatinov91}). Indeed, the general tendency of his result is that the toroidal magnetic
field, which for an axisymmetric mean field corresponds to our $\ov{B}_y$, is pumped towards the equator, 
while the poloidal field (i.e. $\ov{B}_x$ and $\ov{B}_z$) is pumped towards the poles (cf. his Eq. 3.26).

\subsection{Longitudinal pumping effect}
Surprisingly, the longitudinal pumping effect had so far not been identified in simulations of
magnetoconvection, although it is often the strongest pumping effect except if the box
is situated near the pole (Figs. \ref{fig3}--\ref{fig4}, column 2). The predominantly negative sign of $\gamma_y$
signifies that the mean field is advected in the retrograde direction within the bulk of the unstable
layer and the overshoot layer. In most cases, there is a thin layer near the top of the box where the field is 
pumped in the prograde direction. Unlike the vertical pumping effect (see below), here the symmetric part 
of $\alpha$-tensor dominates. From the relations $\gamma^{(x)}_y=\gamma_y-\alpha_{xz}^{\rm S}$ and
$\gamma^{(z)}_y=\gamma_y+\alpha_{xz}^{\rm S}$ (Eq. \ref{up}) it follows that the longitudinal pumping velocities
of $\ov{B}_x$ and $\ov{B}_z$ can be opposite. Since $\alpha_{xz}^{\rm S}$ is positive in the bulk of 
the unstable layer (Fig.~\ref{fig5}, column 3), one finds that $\ov{B}_x$ is pumped mainly in the retrograde 
direction, but $\ov{B}_z$ mainly progradely 
near the top of the unstable layer, while retrogradely near the bottom. This behaviour is reflected in the volume
averages (Fig.~\ref{fig8}). The longitudinal pumping effect is
strongly dependent on latitude; it vanishes at the pole and peaks at the equator. The lowest-order term which
contributes to longitudinal pumping (Krause \& R\"adler \cite{krause80}, p. 193) is 
$\Delta\alpha_{ij}=A g_i\Omega_j+B g_j\Omega_i$, where $A$ and $B$ are true scalars. Clearly, the effect requires 
both gravity and rotation. The antisymmetric part, corresponding to general 
longitudinal pumping, has been computed by R\"udiger \& Kichatinov (\cite{ruediger93}) in the form of 
$\alpha_3 (\Omega_i G_j -  \Omega_j G_i)$, where the coefficient $\alpha_3$ proved to be positive-definite. Here
$\vec{G}$ is the gradient vector of the turbulence intensity, i.e. $\vec{G} =\nabla \ln u_{\rms}$. One obtains
\begin{equation}
\gamma_y \propto \Omega \sin \theta {d\,\ov{u^2}\over dz},
\label{1}
\end{equation}
while for the vertical pumping the simple relation
\begin{equation}
\gamma_z \propto - {d\,\ov{u^2}\over dz}
\label{2}
\end{equation}
results, with only a very weak dependence on the Coriolis number. Thus $\gamma_y$ and $\gamma_z$ should always
satisfy the relation $\gamma_y \gamma_z < 0$, which is indeed the case in all the presented
simulations. According to Eqs (\ref{1}) and (\ref{2}), the nodes which exist in all 
profiles of $\gamma_y$ and $\gamma_z$ should be very close together, which is really true in 
Figs. \ref{fig3}--\ref{fig4}. For completeness we also note that longitudinal pumping cannot
be attributed to the type of rotational anisotropies studied by Kichatinov (\cite{kichatinov91}), since the
resulting pumping effects exist only in the radial and latitudinal directions.

\begin{figure}
\centerline{\psfig{file=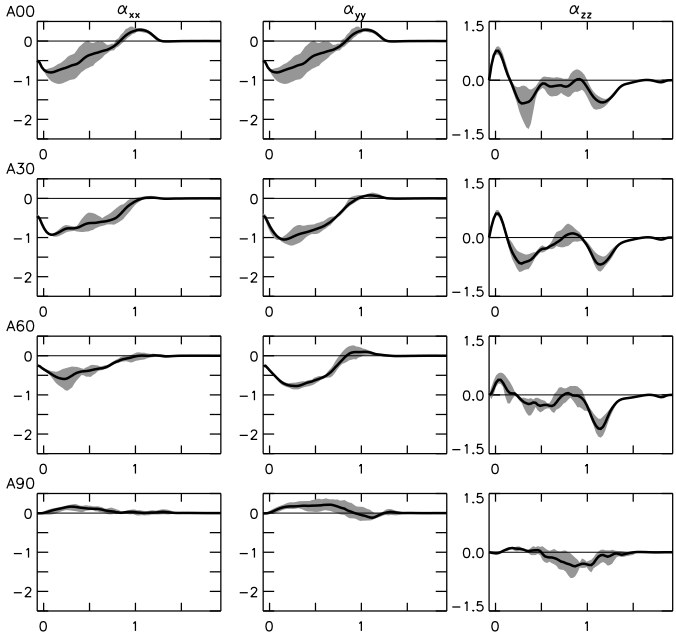,width=9.0cm}}
\caption{Diagonal components of the $\alpha$-tensor (Eq. \protect{\ref{as}}) measured in units of 
$0.01\sqrt{dg}$ as a function of depth in units of $d$ from runs of series A performed at different 
values of $\theta$ (cf. also Fig.~\protect{\ref{fig3}} and Table \ref{partable}). 
On symmetry grounds one knows that $\alpha_{xx}=\alpha_{yy}=\alpha_{zz}=0$
in series A90. This was not imposed, so that these coefficients give an indication of the error 
of the method.}
\label{fig6}
\end{figure}

\subsection{Vertical pumping effect}
The main features of the vertical pumping effect are as follows. The direction of vertical pumping is 
downward ($\gamma_z>0$) in the bulk of the unstable layer and in the overshoot layer below it 
(Figs. \ref{fig3}--\ref{fig4}). Near the top of the box there is a thin layer where the pumping is 
directed upwards ($\gamma_z<0$). The strongest effect is observed in the
laminar cases (D45 and E45), where upward and downward speeds can amount to about $15\%$ of the rms
convective velocity (Table~\ref{partable}). In the more turbulent runs, the
maximum speeds are smaller. Since the top layer where the pumping is upward is less pronounced in the turbulent
runs, the decrease is less apparent in the volume-averaged coefficient (Fig.~\ref{fig8}) because there is
less cancellation. There is little dependence  on rotation (compare A45 with B45 and D45 with E45) or 
latitude (Figs. \ref{fig3} and \ref{fig8}). Also the rms convective velocity 
has a depth dependence that is similar in all runs, namely
a plateau at the top of the domain and below that a monotonic decrease with depth.
This in combination with the robustness of the vertical pumping, in particular its virtual independence of
rotation, suggests an explanation in terms of 'turbulent diamagnetism', i.e. by
transport down the gradient of the rms convective velocity (Eq. \ref{2}). Analytical expressions for the diamagnetic
pumping effect derived under the FOSA assumption were given by Kichatinov \& R\"udiger (\cite{kichatinov92}).

\begin{figure}
\centerline{\psfig{file=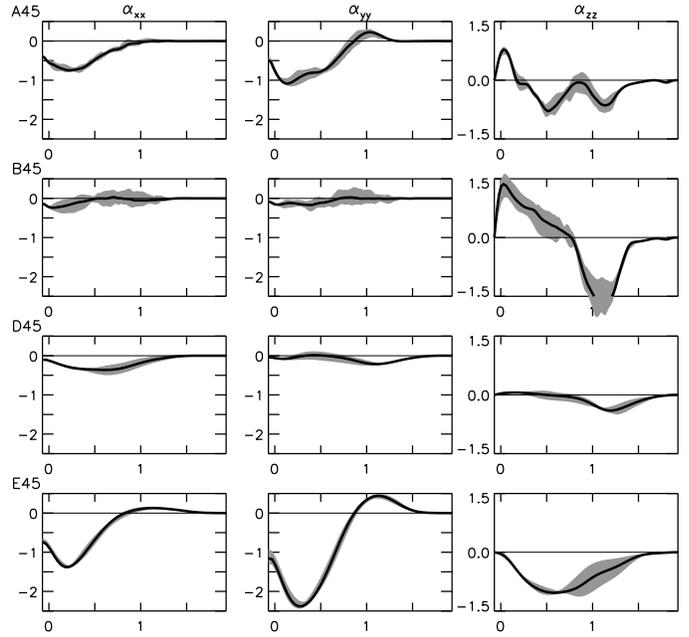,width=9.0cm}}
\caption{Diagonal components of the $\alpha$-tensor (Eq. \protect{\ref{as}}) measured in units of 
$0.01\sqrt{dg}$ as a function of depth in units of $d$ from four different runs performed at
$\theta=45\degr$ (cf. also Fig.~\protect{\ref{fig3}} and Table \ref{partable}).}
\label{fig7}
\end{figure}

If the Rayleigh number is increased from $10^6$ (series B) to $4\cd 10^6$ (series C), so that the convection 
becomes more strongly turbulent, $|\gamma_z|$ decreases. This would appear to be at odds with the result of 
Tobias et al. (\cite{tobias01}), who found more efficient pumping for increased Rayleigh numbers.
But they measure the pumping effect through the ratio of magnetic fluxes in the unstable
layer and the overshoot layer. This ratio is determined both by transport properties and storage
properties of the magnetic field, i.e. by the balance between pumping,
turbulent diffusion in the convection zone, and molecular diffusion in the overshoot layer ($\eta$).
If $\eta$ is decreased, flux concentrations decay more slowly, so that the storage capacity is increased.
Since the increase of the Rayleigh number in Tobias et al. (\cite{tobias01}) is accompanied by a decrease of
$\eta$, the flux ratio increases, even if the pumping is somewhat less efficient.

\begin{figure}
\centerline{\psfig{file=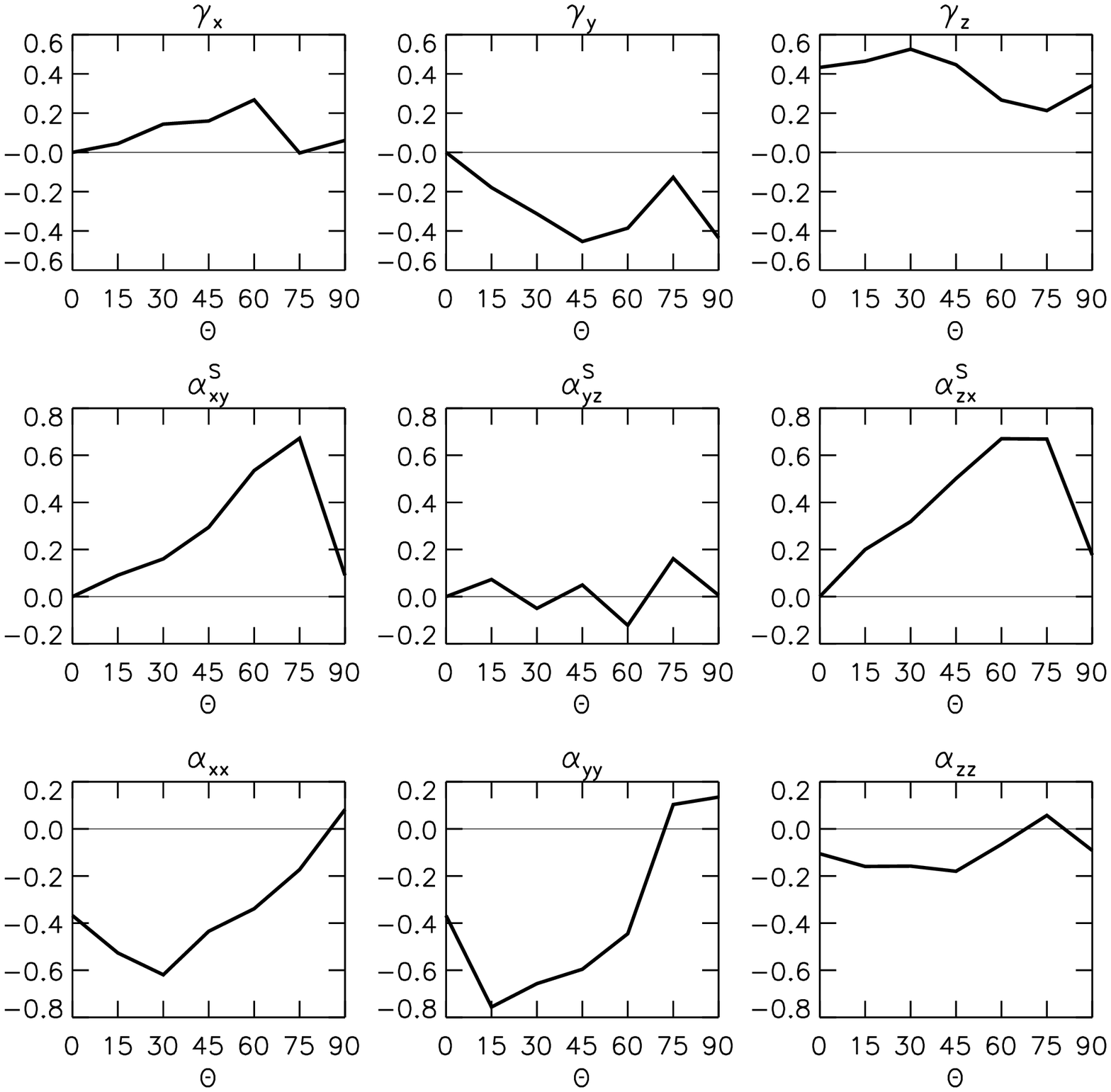,width=9.0cm}}
\centerline{\psfig{file=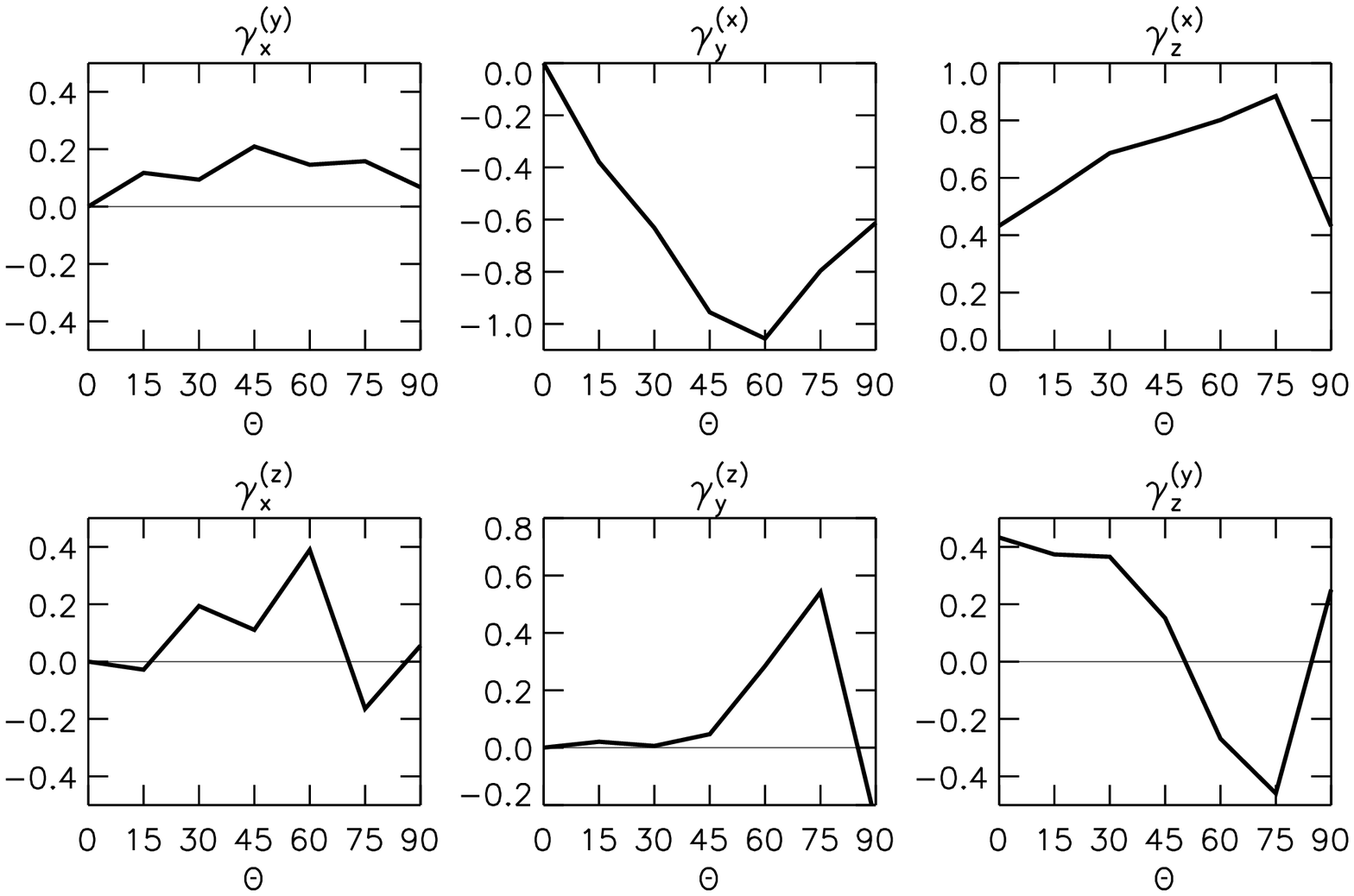,width=9.0cm}}
\caption{Averages over the unstable layer ($0<z<1$) of mean-field transport coefficients for series A
(Figs. \protect{\ref{fig3}}--\protect{\ref{fig7}}), measured in units of $0.01\sqrt{dg}$, as a function of angular distance
to the south pole. The field-direction dependent pumping effects, shown for clarity in the bottom six figures, 
are related to the other coefficients through Eq.~(\protect{\ref{up}}). A rough indication of the 
errors is provided by the shaded areas in 
Figs. \protect{\ref{fig3}}--\protect{\ref{fig7}}; the resulting error in the $z$-averages is typically $0.15$ in units of
$0.01\sqrt{dg}$. Thus the results for $\alpha_{yz}^{\rm S}$ are not significant, while those for $\gamma_x$, 
$\alpha_{zz}$, and $\gamma_x^{(y)}$ are marginally significant.}
\label{fig8}
\end{figure}

In series A it can be seen that the symmetric contribution $\alpha_{xy}^{\rm S}$ is positive and that it 
increases if $\theta$ increases from $15\degr$ to $75\degr$ (Fig.~\ref{fig5}). Using the relations 
$\gamma_z^{(x)}=\gamma_z+\alpha_{xy}^{\rm S}$ and $\gamma_z^{(y)}=\gamma_z-\alpha_{xy}^{\rm S}$ 
(Eq. \ref{up}) one can infer from this that the pumping speeds for $\ov{B}_x$ and $\ov{B}_y$ deviate 
increasingly from one another: the downward pumping of $\ov{B}_x$ becomes faster, and the top layer 
where $\ov{B}_y$ is pumped upward becomes deeper; this is reflected in Fig.~\ref{fig8}.
A comparison of runs B45 with A45 and D45 with E45 shows that
$\alpha_{xy}^{\rm S}$ increases with increasing rotation rate, while $\gamma_z$ is hardly affected
(Figs. \ref{fig3}--\ref{fig5}).
In run D45, which has the weakest rotation of all, the value of $\alpha_{xy}^{\rm S}$ is not
above the noise level, and the same is probably the case for run B45. 
From the symmetry considerations of Krause \& R\"adler (\cite{krause80}), p. 193, it is inferred that the 
lowest-order contribution to $\gamma_z$ stems from the term $\Delta\alpha_{ij}=A\epsilon_{ikj}g_k$, while 
that to $\alpha_{xy}^{\rm S}$ stems from 
$\Delta\alpha_{ij}=A\Omega_i\epsilon_{jkl}g_k\Omega_l+B\Omega_j\epsilon_{ikl}g_k\Omega_l$,
where $A$ and $B$ are true scalars. The latter term also gives rise to latitudinal pumping, as mentioned above. 
Unlike $\gamma_z$, the coefficient $\alpha_{xy}^{\rm S}$ is rotationally induced, and this is
confirmed by the simulations.


\subsection{Alpha effect}
As a byproduct of the determination of the pumping effects, we obtain results for the diagonal part of
the $\alpha$-tensor. These results are an extension to what was reported in Paper I, where all
simulations were done for $\mb{Ra}=50\,000$ and $\theta=0\degr$. The sign and magnitude of $\alpha_{yy}$ are of special 
interest, since this component is responsible for the generation of the large-scale magnetic field in solar-type stars.
The depth dependence of $\alpha_{xx}$ and $\alpha_{yy}$ has a common shape in {\em all} series, provided there is 
sufficient rotation, namely a negative sign in the bulk of the convection zone, and a positive sign in the overshooting layer
(Figs. \ref{fig6}--\ref{fig7}). These features, which are expected for the southern hemisphere, are explained, e.g., in Paper I.
Series A (Fig.~\ref{fig6}) provides detailed results on the latitude dependence. 
The amplitudes of $\alpha_{xx}$ and $\alpha_{yy}$ increase with angular distance from the equator up to a point
close to the south pole, more or less consistent with the commonly assumed $\cos\theta$-function 
(Fig.~\ref{fig8}). At the pole, there appears to be an unexpected dip. 
The amplitudes of $\alpha_{xx}$ and $\alpha_{yy}$ are approximately the same near 
the pole -- for $\theta=0\degr$ they are equal on symmetry grounds -- but closer to the
equator $\alpha_{yy}$ is somewhat larger (especially E45; Fig.~\ref{fig7}). Without rotation, 
the diagonal $\alpha$-terms vanish (the results in Fig.~\ref{fig7} for D45, which has the weakest rotation,
are probably not significantly different from zero), and the increase of $\alpha_{xx}$ and $\alpha_{yy}$ 
with rotation (compare B45 with A45 and D45 with E45 in Fig.~\ref{fig7}) confirms the results of Paper I. 
The same can be said for $\alpha_{zz}$, which for $\mb{Co}\la 2$ has the opposite sign of $\alpha_{xx}$ 
(tentatively visible for B45 in Fig.~\ref{fig7} but clear in Paper I). If rotation increases beyond a certain 
point, $\alpha_{zz}$ as a function of depth developes multiple sign changes (cf. series A), and its amplitude 
falls behind that of $\alpha_{xx}$, even more so if one also averages over depth (Fig.~\ref{fig8}). 
This is the rotational quenching of the vertical alpha effect reported in Paper I.

\section{Discussion}
We have performed numerical simulations of magnetoconvection in a Cartesian box with a rotating frame
of reference which is meant to represent a section from the lower part of a stellar convection zone.
Through an inversion of the relation between the mean magnetic field and the electromotive force,
we have measured pumping effects in all three directions as well as the symmetric elements of the alpha tensor.
For the first time, the existence of non-radial pumping effects is hereby established numerically, and quantified
in terms of gamma coefficients.
Previously, their existence was known only on the basis of symmetry considerations in the framework of mean-field 
electrodynamics (Krause \& R\"adler \cite{krause80}, p. 193), and analytical results based on the first-order 
smoothing approximation (Kichatinov \cite{kichatinov91}; R\"udiger \& Kichatinov \cite{ruediger93}). 
Such analytical results are available only for some of the coefficients that appear in the expansion 
of the EMF given by Krause \& R\"adler (\cite{krause80}). Moreover, the validity of 
results based on FOSA is questionable for real and simulated stellar convection zones alike, and 
they can be readily evaluated only for simple turbulence models but not for actual convective flows. 
For these reasons, it is difficult at present to explain all of the phenomena that are found in the 
simulations, although we did observe that several qualitative features of the expressions derived using
FOSA are confirmed by the simulations, as was the case in Paper I.

The Rayleigh number was varied between $10^4$, corresponding to laminar convection, and $4\cd 10^6$, 
corresponding to mildly turbulent convection. Within this parameter range, we found remarkably consistent 
behaviour of the dynamo coefficients. Clearly, the regime of stellar convection, which is characterized by 
Rayleigh numbers many orders of magnitude larger, remains far out of reach because of the limited spatial 
resolution. The main resulting difference between the simulations and stellar convection is as follows. 
In the lower part of stellar convection zones, the superadiabaticity is very small (typically 
$\delta\approx 10^{-6}$) and, according to the standard mixing-length expressions (e.g., Stix~\cite{stix89}, 
p. 198), this results in velocities of the order of $50$~m~s$^{-1}$, or $\mb{Ma}\approx 10^{-4}$.
In the simulations, convection is less turbulent. After the onset of convection,
the superadiabaticity in the unstable layer decreases from its initial value, $\delta_2=0.1$, to a value 
of at least about $10^{-2}$ (series C). The convective velocities are therefore much larger than in the Sun,
if expressed in m~s$^{-1}$ or as a Mach number. We may illustrate this for series A, where the unstable layer covers
$3.7$ pressure scale heights (Table \ref{partable}). Counting upwards from the bottom of the solar
convection zone, this corresponds to about $90\,000$ km. Setting $g\approx 540$ m~s$^{-2}$, one would obtain
$\sqrt{dg}\approx 2\cd 10^5$~m~s$^{-1}$ for the unit of velocity. Thus the rms convective velocity of
series A (Table \ref{partable}) would correspond to about $2\cd 10^4$~m~s$^{-1}$, i.e. approximately 
$400$ times larger than the solar value. About the same ratio holds for the Mach numbers.
Only simulations with a much higher resolution and much smaller entropy gradients can improve this ratio.
However, the rotation rate was chosen such that the Coriolis number, i.e. the relative importance of the
Coriolis force in the equation of motion, is within the appropriate range.
Also, it is to be expected that beyond some Rayleigh number, a further decrease of the molecular viscosity
or the magnetic diffusivity affects only the smallest scales while having little or no effect on the large
scale. From the consistency of the results one may therefore suspect that the current simulations already
provide some clues about the mean-field transport properties of stellar convection, if properly scaled.

In order to assess the astrophysical relevance of the results, the pumping speeds should therefore be compared with
the rms convective velocity for the unstable layer (Table~\ref{partable}) rather than be directly expressed in
m~s$^{-1}$. The latitudinal pumping effect for the toroidal mean field ($\ov{B}_y$) was found to be mainly
equatorward with speeds up to about $0.1 u_{\rms}$ in the turbulent runs for $\theta=45\degr$, while
the volume averages ($\gamma_x^{(y)}=\gamma_x+\alpha_{yz}^{\rm S}$ in Fig.~\ref{fig8}) are less than about
$0.03 u_{\rms}$. The equatorward motion in the solar butterfly diagram corresponds to a speed of
about $1$~m~s$^{-1}$, so that one may conclude that the latitudinal pumping effect of the toroidal mean
field has the same order of magnitude. We therefore point out the possibility that the equatorward motion
of solar magnetic activity is not the result of a dynamo wave or meridional bulk motion, but of latitudinal
pumping of the toroidal mean magnetic field. Results for the latitudinal pumping of the radial mean field
do not show consistent behaviour; the poleward motion at the top of the box found in some runs
amounts to at most $0.1 u_{\rms}$, but this should be treated with caution because of contamination with higher-order
terms due to the strong gradients in the mean field near the top of the box. 
Since the simulations are not meant to represent the photosphere, the relevance of this result for the observed poleward
motion of poloidal fields at the solar surface is in any case hard to establish. We merely point out the
possibility that the observed motion is the result of pumping of the deep-seated global poloidal field
rather than of meridional bulk surface motion. The longitudinal pumping attains maximal speeds of
about $0.2 u_{\rms}$ in the retrograde direction. Again, the mainly prograde pumping of the poloidal mean field 
near the top of the box should be treated with caution because of the possibility of strong higher-order
contributions. Perhaps, if carried over to the surface of the Sun, it could explain the observed motion of
the active longitudes in the solar wind relative to solar surface rotation 
(Ruzmaikin et al.~\cite{ruzmaikin01}). We also note that preliminary calculations of a mean-field
model indicate that the inclusion of longitudinal pumping in an $\alpha^2$-type dynamo can lead
to oscillating solutions, which are rare for such models. This can be relevant for explaining 
magnetic cycles in stars which do not have differential rotation (R\"udiger et al.~\cite{ruediger02a}).
The downward pumping in the turbulent runs attains speeds of up to about $0.1 u_{\rms}$, and the volume 
average about half of that. In the lower part of the solar convection zone this translates to a few m~s$^{-1}$. Such 
speeds may be sufficient to overcome the magnetic buoyancy of flux generated in the
convection zone and transport the flux downward into the overshoot layer. This layer coincides more or less with the 
tachocline, a region where the rotation rate changes rather suddenly with depth. 
Through stretching of field lines, the shearing motions can generate from poloidal field components a strong  
field of predominantly toroidal orientation, as is required in order to account for various features of the 
large-scale solar magnetic field.

The origin of the poloidal magnetic field and its locus of generation still remain an open question.
According to one likely scenario, the interface dynamo (Parker \cite{parker93},
Ossendrijver \& Hoyng \cite{ossendrijver97}, Charbonneau \& MacGregor \cite{charbonneau97}), the poloidal
field is generated by an alpha effect in the lower half of the convection zone.
Our results for the coefficient $\alpha_{yy}$, which corresponds to $\alpha_{\phi\phi}$ in spherical
coordinates, are of particular interest for interface-type stellar dynamo models. We should stress however that it
cannot be decided from the present simulations whether the large-scale magnetic field of solar-type 
stars is generated by the convective alpha effect investigated here or by a different alpha effect 
such as that resulting from the buoyancy instability of magnetic flux tubes in the overshoot layer. 
We confirm that the convective alpha effect is strongest near the poles. For the Sun,
this should, in combination with the differential rotation near the poles, as inferred from 
helioseismic inversions, result in strong dynamo action near the poles 
(R\"udiger \& Brandenburg~\cite{ruediger95}), which is not observed -- 
sunspots are concentrated in belts that extend to a latitude of at most $40\degr$ on either side of the equator.
This well-known problem may perhaps be resolved by taking into account meridional circulation, and by noting that
the buoyancy instability of toroidal magnetic flux tubes anchored in the overshoot layer is reduced or 
suppressed at high latitudes (Ferriz-Mas \& Sch\"ussler~\cite{ferrizmas95}).
The sign of $\alpha_{yy}$ confirms what is generally found in MHD simulations (e.g. Paper I), namely a negative
sign in the bulk of the convection zone on the southern hemisphere, i.e. a positive sign on the northern 
hemisphere, and a sign change at the bottom of the convection zone. This distribution of signs
puts a tough constraint on theories of the solar dynamo. Generally, mean-field dynamo models that
incorporate solar differential rotation, in particular a positive $d\Omega/dr$ near the equator near the
base of the convection zone, produce a butterfly diagram with the required equatorward migration of
the magnetic activity belts only in combination with a negative $\alpha_{\phi\phi}$ on the northern hemisphere.
There are various ways in which this dilemma could be resolved; perhaps the relevant dynamo effect is
located at the very bottom of the convection zone where the sign has changed. If $\alpha_{\phi\phi}$
is positive on the northern hemisphere, equatorward migration can be obtained if a suitable meridional 
circulation is included (Choudhuri et al.~\cite{choudhuri95}; Durney~\cite{durney96}; 
K\"uker et al.~\cite{kueker01}; R\"udiger et al.~\cite{ruediger01}; R\"udiger \& Arlt~\cite{ruediger02}).
As mentioned above, the present results suggest the alternative possibility of latitudinal pumping.
This suggestion needs however to be verified in mean-field models that take into account all relevant
dynamo coefficients as well as solar differential rotation.

\begin{acknowledgements}
M.O. acknowledges financial support by the Deutsche Forschungs\-gemein\-schaft. The calculations were
carried out largely on the Origin 2000 of the Rechenzentrum of the University of Freiburg i. Br..
\end{acknowledgements}

\end{document}